\documentclass[epsfig]{mn2e}
\usepackage{times,graphics,epsfig}
\newif\ifAMStwofonts

\title{Radio observations of NGC~2808 and other globular clusters: constraints on intermediate mass black holes}

\author[Maccarone et al.] {Thomas J. Maccarone\\ University of
Southampton, School of Physics and Astronomy, Highfield Campus,
Southampton, Hampshire, SO17 1BJ \newauthor Mathieu Servillat\\ Centre
d'\'Etude Spatiale des Rayonnments, Universit\'e Paul Sabatier, CNRS -
9 avenue du Colonel Roche, 31400 Toulouse, France}

\date{}

\begin{document}

\maketitle

\label{firstpage}

\def\simlt{\mathrel{\rlap{\lower 3pt\hbox{$\sim$}}
        \raise 2.0pt\hbox{$<$}}}
\def\simgt{\mathrel{\rlap{\lower 3pt\hbox{$\sim$}}
        \raise 2.0pt\hbox{$>$}}}

\input epsf

\begin{abstract}
We present the results of a deep radio observation of the globular
cluster NGC~2808.  We show that there are no sources detected within
the core of the cluster, placing constraints on both the pulsar
population of the cluster and the mass of a possible intermediate mass
black hole in NGC~2808.  We compare the results for this cluster with
other constraints on intermediate mass black holes derived from
accretion measures.  With the exception of G1 in M~31 which has
previously shown radio emission, even with considerably more
conservative assumptions, only the clusters with the poorest of
observational constraints are consistent with falling on the
$M_{BH}-\sigma$ relation.  This result is interpreted in terms of the
fundamental differences between galaxies and globular clusters.
\end{abstract}


\section{Introduction}

Starting with the discoveries of X-ray sources in globular clusters in
the mid-1970's (Clark 1975; Clark, Markert \& Li 1975), considerable
debate took place about whether globular clusters contain black holes
of intermediate masses (i.e. greater than the $\sim$ 20 $M_\odot$
maximum mass for black holes formed through normal single star
evolution, but less than the $10^6 M_\odot$ masses seen in the
smallest galactic nuclei).  It was proposed that the X-ray emission
seen from these clusters was due to accretion onto the central black
holes of material released into the intracluster medium by stellar
mass loss (Bahcall \& Ostriker 1975; Silk \& Arons 1975).
Additionally, increases in the clusters' central velocity dispersions
were seen without corresponding increases in the central optical
luminosity density (Newell, Dacosta \& Norris 1976).  Both arguments
for intermediate mass black holes were refuted in relatively short
order; Type I X-ray bursts were seen from the globular cluster X-ray
sources (Grindlay et al. 1976), and explained to be due to
thermonuclear burning on the surfaces of neutron stars (Woosley \&
Taam 1976).  With regards to the second argument, it was shown that
increases in central mass-to-light ratios of globular clusters were
expected due to mass segregation in the clusters, which leaves the
heavy, dark white dwarfs and neutron stars preferentially in the
cluster cores (Illingworth \& King 1977).

This controversy has been re-kindled recently based on the same lines
of evidence.  An intermediate mass black hole in M~15 has been claimed
(Gerssen et al. 2002) and refuted (Baumgardt et al. 2003) on largely
the same grounds as the debate over the same cluster in the late
1970's.  In fact, it has been argued that the finite number of stars
within the sphere of influence of an intermediate mass black hole in a
globular cluster will make it impossible to demonstrate conclusively
through stellar dynamical measurements made in integrated light that a
cluster contains an intermediate mass black hole (Drukier \& Bailyn
2002).  More indirect proofs may still be feasible, through, for
example, measurements of high velocity stars (Drukier \& Bailyn 2002;
Baumgardt, Gualandris \& Portegies Zwart 2006) or the presence of
unusual binaries which can be best explained by recoil off a black
hole-black hole binary (Colpi, Possenti \& Gualandris 2002).

The inability of stellar dynamics to make clean measurements has led
to the search for accretion constraints on the presence of
intermediate mass black holes.  The first attempts placed upper limits
on the masses of black holes in M15 and 47 Tuc (Grindlay et al. 2001;
Ho et al. 2003), based on particular accretion models.  It was later
pointed out that, following from the fundamental plane relation for
black hole activity (Merloni, Heinz \& Di Matteo 2003; Falcke,
K\"ording \& Markoff 2004), the radio emission expected from an
intermediate mass black hole would be orders of magnitude higher than
that expected from a stellar mass black hole or a neutron star at the
same X-ray luminosity.  This makes radio searches more sensitive, and
also gives the possibility to place at least a crude constraint on
black hole mass based on detection of radio emission (Maccarone 2004).

One piece of strong evidence has emerged for an intermediate mass
black hole in G1, an enigmatic star cluster in M~31 thought by some to
be a globular cluster, but by others to be the core of a stripped
dwarf galaxy, given that its mass is far above that of any globular
cluster in the Milky Way, and that it contains multiple stellar
populations.  Dynamical evidence in G1 suggested the presence of a
20\,000 solar mass black hole (Gebhardt, Rich \& Ho 2005), but was
significant at less than the $3\sigma$ level and based on only a few
pixels worth of HST data.  X-ray measurements of this cluster revealed
a source at $2\times10^{36}$ ergs sec$^{-1}$ (Trudolyubov \&
Priedhorsky 2004), and were suggested to be possible evidence of an
intermediate mass black hole, but was certainly still consistent with
emission from a single X-ray binary or a collection of X-ray binaries
(Pooley \& Rappaport 2005).  The radio flux from G1 was then measured
to be 28 $\mu$Jy (Ulvestad, Greene \& Ho 2007), in good agreement with
the prediction of Maccarone \& K\"ording (2006) that the radio flux
should be 30 $\mu$Jy for the suggested black hole mass and observed
X-ray luminosity.  We note that much of the information from this
introduction is drawn from a recent review by Maccarone \& Knigge
(2007).  For a more general and thorough, but somewhat dated review of
binaries in globular clusters, see Hut et al. (1992).  For a recent
review on the topic of black holes and neutron stars in globular
clusters, see Rasio et al. (2007).

On the other hand, searches for radio emission from globular clusters
thought to be good candidates for hosting intermediate mass black
holes have generally yielded only upper limits (Maccarone, Fender \&
Tzioumis 2005; De Rijcke et al. 2006; Bash et al. 2008).  It is thus
clear that more detailed searches for radio sources from Galactic
globular clusters should be attempted.  In this paper, we report an
upper limit on the radio emission from the globular cluster NGC~2808.
We discuss this result in the context of other radio upper limits from
globular clusters, and show that there are already enough data to cast
severe doubt on suggestions that globular clusters may follow the same
$M_{BH}-\sigma$ relation as galaxies.  

\section{Data Used}

The Australia Telescope Compact array observed NGC~2808 for 12 hours
in array configuration 6D, on 24 January 1992.  The calibrators used
are 0823-500 and 1934-638.  The data were taken at 1.408 and 1.708
GHz, with 64 channels of 2 MHz each each frequency range.  The 1.708
GHz data were especially noisy and were not examined.  We reduced the
1.408 GHz data with MIRIAD (Sault, Teuben \& Wright 1995).  We needed
to excise the four lowest frequency channels, the four highest
frequency channels, and the eight central channels in order to remove
an artefact in the central pixel and considerable noise.  A few other
epochs of radio frequency interference activity were excised, as was
the baseline between antennae 4 and 5 which showed strong noise
throughout the observation.  As a result, the rms obtained in these
data is not as good as one might hope for, but after CLEANing, the
data still produce a noise level of 54 $\mu$Jy, very similar to the
theoretical noise level expected from the included baselines.

\section{Other sources}

Several radio sources are detected in these data, within the tidal
radius of the cluster but outside the half-light radius.  Given that
any radio source in a globular cluster is likely to be a source which
is formed dynamically, which is far heavier than a typical star in the
cluster, or both, any radio source associated with the cluster is
likely to be near the centre of the cluster where dynamical
interactions are most common and where mass segregation leaves the
heaviest objects.  These sources are therefore most likely to be
background active galactic nuclei, although, owing to the relatively
low Galactic lattitude of this cluster ($b=-11.3$ degrees) some may be
foreground object, such as HII regions.  Correlations with X-ray
detections from Servillat, Webb \& Barret (2008) will be discussed in
future work.

\section{Discussion}

A few key points result from the non-detection of radio sources
associated with NGC~2808.  First, the non-detection of any sources in
the core implies that there are no bright radio pulsars here.  The
collision rate, $\Gamma$ in NGC~2808, approximated as
$\Gamma=\rho_c^{3/2}r_c^2$ (Verbunt \& Hut 1987), where $\rho_c$ is
the stellar density in the core, and $r_c$ is the core radius, is
slightly higher than that in 47~Tuc, which has more than 20 known
pulsars (Camilo et al. 2000); pulsar production is expected to be
well-correlated with the collision rate, in the same way that X-ray
source production is enhanced (Gendre et al. 2003; Pooley et al. 2003,
2006).  While none of the pulsars in 47~Tuc is bright enough to be
seen in this exposure at the distance of NGC~2808, the integrated
emission from pulsars in 47~Tuc would be detectable at this distance.
Given that the core of NGC~2808 is only about 4$\times$4 angular
resolution elements for the ATCA's angular resolution at 1.4 GHz, and
that most of the ATCA baselines are sufficiently short that the full
cluster core is well-probed by the baselines, it seems unlikely that
the pulsars are over-resolved by the array.  The level of emission
really is likely to be lower than in 47 Tuc.  Part of the reason for
this may be the lower metallicity of NGC~2808 relative to 47~Tuc -- it
has recently become well established that the metallicity of a
globular cluster affects its probability of hosting an X-ray source
(see Silk \& Arons 1975 for the first suggestion of this effect;
Kundu, Maccarone \& Zepf 2002 for the first definitive evidence; Kim
et al. 2006; Kundu et al. 2007; Sivakoff et al. 2007 for
demonstrations from large samples of elliptical galaxies).  Deeper
observations in radio continuum would be useful for constraining
whether there are real differences in the pulsar properties of these
two clusters, or if the issue is just the combination of deeper data
on 47~Tuc so far, combined with its smaller dispersion measure which
makes pulsar timing detections at low frequencies easier to make.

Second, we consider the case of a possible intermediate mass black
hole in NGC~2808.  The interpretation of the non-detection depends
strongly on the assumptions made about the efficiency of Bondi-Hoyle
accretion (i.e. what fraction of the classical Bondi rate is actually
accreted), the gas density in the cluster, the radiative efficiency of
accretion, and the correlation between X-ray and radio power.  We will
present here both the most likely parameter values and a more
conservative estimate which consists of considering values for each
parameter near the lower end of the plausible range.

A variety of approaches can be used to estimate the fraction of the
Bondi rate at which spherical accretion takes place in Nature.
Pellegrini (2005) examines a sample of active galactic nuclei in
elliptical galaxies, where the gas density can be estimated from X-ray
imaging.  Nearly all these galaxies are underluminous compared to what
would be expected for radiatively efficient accretion at the Bondi
rate.  When radiatively inefficient accretion, with $L_X \propto
\dot{m}^2$ is implemented (as we do here -- see below), the accretion
rates predicted from the Bondi relation are still too low to match the
observed data, but typically by a factor of only about 10-30 -- see
Figure 3 of Pellegrini (2005), with the most underluminous AGN only a
factor of about 1000 below its Bondi rate for radiatively inefficient
accretion.  Various authors have also considered the kinetic power of
jets from AGN in clusters of galaxies based on the bubbles they
inflate in the intracluster medium (see e.g. Dunn, Fabian \& Celotti
2005).  They typically find that at least 10\% of the power which
would be generated by efficient accretion at the Bondi rate is
required (RJ Dunn, private communication).  Finally, Perna et
al. (2003) have estimated that the lack of detection of isolated
neutron stars accreting from the interstellar medium implies a
fraction of Bondi accretion of $10^{-2}-10^{-3}$.  We thus take
$3\times10^{-2}$ times the Bondi rate as the most likely value and
$10^{-3}$ as our more conservative reasonable value.

The gas content of globular clusters is also an issue of debate.  Four
globular clusters clearly show evidence for an intracluster medium,
with the rest generally showing upper limits consistent with being
well above the measured values in the clusters with good measurements
(e.g. Bowyer et al. 2007; Knapp et al. 1996).  The two Galactic
globular clusters with evidence for gas, 47~Tuc and M~15 show
systematic correlations between the dispersion measures of their
pulsars and the accelerations of their pulsars, indicating that the
pulses from the pulsars in the back of the cluster are moving through
larger column densities of material (Freire et al. 2001;2003).  The
cluster G1 in M31 shows evidence for gas in that it verifies the
picture we are presenting here.  Finally, the cluster RZ2109 shows a
strong, broad [O III] emission line, powered by the accreting, likely
stellar mass, black hole within it (Maccarone et al. 2007; Zepf et
al. 2007).

Gas densities can also be estimated based on empirical knowledge about
stellar mass loss.  Pfahl \& Rappaport (2001) showed that if the
gravitational potentials of globular clusters are ignored, so that the
mass loss from stars is treated as free expansion, that core densities
of gas of $n_H \approx 1 \left(\frac{M_c}{10^5
M_\odot}\right)\left(\frac{v_w}{20 \rm{km
sec^{-1}}}\right)^{-1}\left(\frac{r_c}{0.5\rm{pc}}\right)^{-2}$
cm$^{-3}$, would result, where $M_c$ is the cluster core mass, $r_c$
is the cluster core radius, and $v_w$ is the characteristic wind
velocity for the outflows.  When we apply this result within this
paper, we will assume a characteristic wind speed of 50 km sec$^{-1}$
(allowing for a rather conservative limit), a cluster mass-to-light
ratio of 2 (e.g. Piatek et al. 1994), and will take the central
luminosity densities and core radii from the Harris (1996) catalog.
We note that the values inferred from this equation are a factor of
about 2 higher than the values derived from pulsar measurements, but
that the pulsar measurements are sensitive only to ionized gas.

A conservative estimate of the gas content in NGC~2808 would be that
it is similar to the gas content in 47~Tuc (i.e. $n_H$=0.07
H cm$^{-3}$).  In fact, it is unlikely that the gas content in NGC~2808
will be as low as in 47~Tuc.  Pure stellar wind mass loss should yield
a gas density of 0.5 H cm$^{-3}$, without any retention of the gas in
both these clusters -- it is likely that millisecond pulsar winds are
responsible for removing some of the gas in 47~Tuc.  Faulkner et
al. (1991) found tentative evidence for 200 $M_\odot$ of neutral
hydrogen in NGC~2808, presuming the gas is located predominantly
in the cluster core.  If this were confirmed, it would represent a gas
density about 4 orders of magnitude higher than the values used in
this paper.  However, this gas density is highly unlikely - an
unusually low dust-to-gas ratio would be required in order not to see
reddening within the cluster, since $n_H$=5\,500 H cm$^{-3}$ with a
cluster core radius of 0.7 pc gives a column density $N_H$ of more
than $10^{22}$ H cm$^{-2}$.  Furthermore, if there really were
this a gas density in the core of NGC~2808 of 5500 H cm$^{-3}$, not
only an intermediate mass black hole, but also the isolated neutron
stars in the cluster would be easily detectable as X-ray sources (see
e.g. Pfahl \& Rappaport 2001).

The radiative efficiency of accretion and the relation between radio
and X-ray powers have, in fact, become well established in recent
years.  The fundamental plane relation for black hole activity, $L_R
\propto L_X^{0.6}M_{BH}^{0.8}$ (Merloni et al. 2003) requires
radiatively inefficient accretion (e.g. from an advection dominated
accretion flow -- see Narayan \& Yi 1995) in the low/hard state
(i.e. below about 2\% of the Eddington limit -- Maccarone 2003;
Maccarone, Gallo \& Fender 2003).  We thus assume that the radiative
efficiency of accretion is $0.1~c^2$ for $L>0.02~L_{EDD}$ and
0.5~$\dot{m}c^4/L_{EDD}$ for lower luminosities, so that the function
is continuous at $L=0.02 L_{EDD}$.  The different radio/X-ray
relations for black holes and neutron stars (Migliari \& Fender 2006)
argue strongly for a radiative efficiency law where $L_X \propto
\dot{m}^2$ in the low hard state of black holes, and the lack of
abrupt changes in luminosity at the state transitions to and from the
high soft state, which is widely believed to be a radiatively
efficient standard optically thick, geometrically thin accretion disk
(Shakura \& Sunyaev 1973) argue for a smooth transition in radiative
efficiency (Maccarone 2005; Russell et al. 2007).  The mass term in
the fundamental plane relation follows in a straightforward manner
from standard synchrotron theory for compact conical jets (Blandford
\& K\"onigl 1979; Falcke \& Biermann 1995; Heinz \& Sunayev 2002).
Confidence in this relation is also bolstered by the fact that the
scatter in the correlations reduces to being consistent with
measurement errors when AGN samples are chosen carefully, so as to
include only AGN in the low/hard state and without significant
relativistic beaming (K\"ording, Falcke \& Corbel 2006).  Scatter
nonetheless, may exist -- the most likely reasons being due to
variations in black hole spins, and Doppler boosting (although there
are several lines of indirect evidence that the jets from low
luminosity systems are only mildly relativistic -- see e.g. Gallo,
Fender \& Pooley 2003).  Below we describe how its effects may be
parameterized.

We then can obtain two limits on the mass of the black hole based on
the radio data.  In all cases, we use the radiative efficiency
prescription described above, and the value for the gas density given
by Pfahl \& Rappaport (2001).  The conservative limit on black hole
mass will come from taking the accretion rate to be $10^{-3}$ of the
Bondi rate, and using the $5\sigma$ upper limit for the radio flux.
Furthermore, in the two clusters where there are estimates of the
ionized gas mass from pulsar velocity dispersions, this gas density,
rather than the density estimated from Pfahl \& Rappaport (2001) is
used for the conservative limit.  This can be regarded as a clear
upper limit on the mass of a black hole.  The second limit can be
obtained by using the $3\sigma$ upper limit on the radio flux and
using 3\% of the Bondi rate as the most likely accretion rate.  For
NGC~2808, the conservative upper limit on the radio flux density will
then be 270 $\mu$Jy, while the more realistic upper limit is 162
$\mu$Jy.  These yield limits on the black hole mass from the data of
2100 $M_\odot$ and 370 $M_\odot$, respectively.

For the reader who wishes to consider cases which span an even larger
range of parameter space than that covered by our best guess and our
more conservative estimate, we derive the scaling relations for mass
versus other parameter values.  Let us define $b$ to be the fraction
of the Bondi rate one assumes, and $p$ to be the fraction of the
fundamental plane radio luminosity used and $g$ to be the fraction of
the Pfahl \& Rappaport gas density used.  Then, since $\dot{m} \propto
bg M^2$; $L_X \propto \dot{m}^2$; and $L_R \propto b L_X^{0.6}
M^{0.8}$, then one can see that $L_R \propto p (bg)^{1.2} M^{3.2}$.
Solving for $M$, instead, one finds that $M \propto p^{0.31}
(bg)^{0.38}$ -- the errors in the mass inferred are a factor of
roughly 2 even if there exists an order of magnitude error in the gas
density, the fraction of the Bondi rate used, or the scaling from
X-rays to radio provided by the fundamental plane relation.  Since we
have taken our more conservative case to be the multiplications of a
conservative estimate of each parameter value, it is very unlikely
that our conservative upper limits on black hole masses will be too
low by even an order of magnitude.

We can contrast these results with previous discussion of a possible
intermediate mass black hole in NGC~2808.  NGC~2808 is considered a
good candidate to host an intermediate mass black hole on the basis of
its morphological properties.  It has a large ratio of core radius to
half-light radius -- expected since an intermediate mass black hole
will heat its host cluster dynamically (Trenti 2006).  It also shows a
density cusp (Noyola \& Gebhardt 2006) argued to form a piece of the
evidence for a black hole between 110 and 3100 solar masses (Miocchi
2007), assuming that the cusp is Bahcall-Wolf (1975) cusp, which
arises from the deeper gravitational potential well near an IMBH than
would otherwise be expected in a globular cluster.  On the basis of a
lack of X-ray sources down to $10^{32}$ ergs/sec at the position of
the cluster's centre, Servillat et al. (2008) suggested an upper limit
of about 290 $M_\odot$ on the mass of an intermediate mass black hole
in the cluster.  The assumptions used by Servillat et al. (2008) are
similar to our most reasonable case.  In this case, the radio and
X-ray provide rather similar limits on the black hole mass -- in cases
where the gas density is relatively high, as is the most reasonable
assumption for NGC~2808, X-ray measurements and radio measurements are
about equally sensitive.  Only when either the gas density, or the
fraction of the Bondi rate at which accretion actually takes place are
low, does the advantage of using radio data really manifest itself.

\section{Comparison with other clusters}

Numerous groups have now made observations of globular clusters in the
radio with the aims of testing whether they contain intermediate mass
black holes.  There have been many different sets of parameters used
for converting the observations into constraints on the masses of the
black holes in the cluster.  We collect observed constraints on radio
flux densities, along with the conservative and most likely
constraints in Table 1.

\begin{table*}
\caption{Summary of radio continuum observations of globular clusters.
Where upper limits are given, they are at the 3$\sigma$ level.  It
should be noted that there does exist a 2$\sigma$ detection of a radio
source in NGC~6266 (Bash et al. 2008).  All distances come from Harris
(1996), except the distance to G1 which comes from Holland (1998).
Mass constraints from radio emission are given using the preferred and
conservative set of parameters from above.  The values of $n_H$ listed
in the table are those expected using the relation of Pfahl \&
Rappaport (2001) for the cluster parameter values in the Harris
catalog.  These are used generically, unless another value is quoted
in the literature; if the literature value is larger, it is used in
both cases.  If the literature value is smaller, it is used as the
conservative value, while the PR value is used as the most likely one,
since the literature values probe only one phase of the ISM.  The
dynamical mass estimate for G1 comes from Gebhardt, Rich \& Ho (2005);
for $\omega$~Cen comes from Noyola et al. (2008); for 47~Tuc comes
from McLaughlin et al. (2006).  The others come from the $M-\sigma$
relation -- with the values for M~15, M~30 and M~62 taken from Bash et
al., and the remainder computed using the relation of Tremaine et
al. (2002).  Values of $\sigma$ are taken from Pryor \& Meylan (1993),
with the exceptions of Pal~2 and NGC~6440 which are taken from
photometric modeling of Gnedin et al. (2002) -- which yields velocity
dispersions of 7 km/sec and 20 km/sec, respectively.  The radio data
are taken from: Ulvestad et al. (2007) for G1; Maccarone, Fender \&
Tzioumis (2005) for $\omega$~Cen, Bash et al. (2007) for M~15, M~60
and NGC~6266; De Rijcke et al. (2006) for NGC~6397 and 47~Tuc; Knapp
et al. (1996) for Pal 2, NGC~1851, NGC~6440 and NGC~7099; and are new
to this paper for NGC~2808.  Knapp et al. (1996) also presented
observations of NGC~6624, but the presence of a bright X-ray binary,
4U~1820-30 within 1'' of the core of this cluster makes it difficult
to estimate the radio flux from any possible IMBH in this cluster.
The first listed radio mass limit is the more conservative one, while
the second is the more likely one.}
\label{summarytable}
\begin{center}
\begin{tabular}{llllllll}\hline
Cluster& Distance (kpc) & $L_R$ ($\mu$Jy kpc$^2$)&$M_{BH,dyn}$ &$M_{BH,rad}$&$n_H$ (cm$^{-3})$\\ \hline
G1& 780&1.7$\times10^7$&18000&4500&$\sim1$\\
$\omega$ Cen&5.3&$<$ 2700&4$\times10^{4}$&2340/390&0.044\\
M~15&10.3&$<2700$&1000&1150/140&0.42/0.2\\
NGC~6397&2.7&$<1600$& 50 &1000/170&0.16\\
47~Tuc&4.5&$<4600$&900$\pm$900&2300/200&0.28/0.07\\
M~80&10.0&$<3600$&1600&1250/210&0.21\\
NGC~6266&6.9&$<$ 1700&3000&900/160&0.41\\
NGC~2808&9.5&$<14600$& 2700 &2100/370&0.24\\
Pal~2&27.6&$<$ 34300& 202 &4400/750&0.09\\
NGC~1851&12.1&$<11900$& 1000 &1500/270&0.37\\
NGC~6440&8.4&$<4900$& 13000 &925/192&0.51\\
NGC~7099&8.0&$<5200$& 70 &1800/300&0.13\\
\end{tabular}
\end{center}
\end{table*}

Only G1 is detected in the radio.  Many of the existing upper limits,
even with the conservative parameter values taken, are well below the
predictions made from assuming that the clusters fit on the $M-\sigma$
relation.  This should not be surprising.  There is no physical reason
why the $M-\sigma$ relation should apply to globular clusters.
Globular clusters fit to King models, while other classes of objects
fit to De Vaucouleurs or Sersic models.  While both classes of objects
do show some mild deviations from these simple parameterisations of
surface brightness, it has been shown clearly that the homology of
these different classes of objects is quite different.  Again, this is
not surprising -- globular clusters are dynamically relaxed and show
little or no evidence of containing dark matter, while the galaxies
which have similar values of $\sigma$, the dwarf spheroidal galaxies,
are among the most dark matter dominated structures in the universe.
While it is true that the locus of globular clusters is in a region of
parameter space that intersects the fundamental plane relation for
bulges (Dressler et al. 1987; Djorgovski \& Davis 1987) and the
fundamental manifold relation which is extended also to include
clusters of galaxies (Zaritsky, Gonzalez \& Zabludoff 2006), the slope
for the globular clusters in nearly any two dimensional projection is
quite different from the slope for the dark matter dominated systems
(e.g. Burstein et al. 1997).

Some special attention should be devoted to the case of $\omega$~Cen,
since it presents the strongest differences between claims from
stellar dynamics measurements and accretion constraints.  On its face,
it appears to present a clear discrepancy between the predictions of
an accretion based model, and the results from a stellar dynamical
estimate which is far more sophisticated than a mere application of
the $M-\sigma$ relation -- in this cluster.  Noyola et al. (2008) have
presented a rotation curve for the cluster indicating that the best
fitting mass of a central dark object is $4\times10^4 M_\odot$.  Even
using our more conservative formulation, a black hole of $4\times10^4
M_\odot$ in $\omega$~Cen would yield an X-ray luminosity of
$3\times10^{34}$ ergs s$^{-1}$ -- two orders of magnitude higher than the
observational upper limit from Gendre, Barret \& Webb (2003) -- and a
radio flux of about 250 mJy -- more than enough to have called
attention to itself long before the $M-\sigma$ relation had even first
been presented.  One would need to scale the accretion rate downwards
by another factor of 1000 from these more conservative parameter
values in order for the accretion from the black hole not to be
detectable.  The cause of the discrepancy is thus almost certainly
with the dynamical mass estimate.  However, since an axisymmetric
orbit-based model indicates that a black hole is significant at
slightly less than the 2$\sigma$ level (Noyola et al. 2008), the issue
may simply be random measurement errors, with no methodological
problems in either the accretion-based or dynamics-based measurements.
Additionally, single-epoch spectroscopy of integrated light leaves in
additional radial velocity components associated with binary motions,
which can be a significant source of excess radial velocity for
clusters of relatively low central density such as $\omega$~Cen (see
e.g. Kouwenhoven \& De Grijs 2008).  Therefore, we find the radio and
X-ray constraints on the black hole mass in $\omega$~Cen to be
indicative of genuine upper limits on the black hole mass which argue
strongly against a black hole of $\sim10^4 M_\odot$, but which are not
necessarily inconsistent with the results from dynamical studies.

\section{Acknowledgments}
TJM is grateful to Rob Fender for useful advice about data analysis,
and Anthony Gonzalez, Dennis Zaritsky and Dean McLaughlin for useful
conversations and correspondence regarding the fundamental planes and
manifolds for globular clusters, galaxies and clusters of galaxies.
MS thanks the University of Southampton for hospitality where a
portion of this work was done.  We both thank Natalie Webb for a
review of the manuscript prior to publication.

\label{lastpage}
\end{document}